 \documentclass[preprint,5p,times,twocolumn,sort&compress]{elsarticle}

\usepackage{amssymb}
\usepackage{bm}

\usepackage[utf8]{inputenc} 
\usepackage{booktabs,amsmath}
\usepackage{bm,amsmath,amssymb,amsfonts,latexsym,psfrag}
\usepackage{amssymb}
 \usepackage{array}
\usepackage{mathtools}
\usepackage{amsthm}
\usepackage{newlfont}
\usepackage{esdiff}
\usepackage{multirow}
\usepackage{graphics}
\usepackage{graphicx}
\usepackage{color}
\usepackage{tabularx}
\usepackage{bm}
\usepackage{natbib}
\usepackage{url}
\usepackage{bm}
\usepackage[colorlinks=true,citecolor=cyan,urlcolor=blue,bookmarks=true,bookmarks=true,bookmarksopen=true,bookmarksnumbered=true,bookmarksopenlevel=3]{hyperref}

\definecolor{airforceblue}{rgb}{0.36, 0.54, 0.66}
\definecolor{steelblue}{rgb}{0.27, 0.51, 0.71}
\definecolor{amber}{rgb}{1.0, 0.49, 0.0}

\journal{Physics Letters B}

\begin{document}

\begin{frontmatter}



\title{ {\bf Perturbative unitarity bounds for effective composite models}
\tnoteref{doi_1}
}
\tnotetext[doi_1]{The present paper is the merged version of original article and Erratum. DOI of original article: \href{https://doi.org/10.1016/j.physletb.2019.06.042}{https://doi.org/10.1016/j.physletb.2019.06.042}\\ DOI of Erratum: \href{https://doi.org/10.1016/j.physletb.2019.134990}{https://doi.org/10.1016/j.physletb.2019.134990}}

\author[label1]{{\scshape S.~Biondini}}
\address[label1]{Van Swinderen Institute, University of Groningen,
Nijenborgh 4, NL-9747 AG Groningen, Netherlands}

\author[label2]{{\scshape R.~Leonardi}}
\address[label2]{Istituto Nazionale di Fisica Nucleare, Sezione di Perugia, Via A.~Pascoli, I-06123 Perugia, Italy}
\author[label2]{{\scshape O.~Panella}}
\author[label3,label4]{{\scshape M.~Presilla}}
\address[label3]{Dipartimento di Fisica e Astronomia ”Galileo Galielei”,
Universit\`{a} degli Studi di Padova, 
Via Marzolo, I-35131, Padova, Italy}
\address[label4]{Istituto Nazionale di Fisica Nucleare, Sezione di Padova, 
Via Marzolo, I-35131, Padova, Italy}


\begin{abstract}
In this paper we present the partial wave unitarity bound in the parameter space of dimension-5 and dimension-6 effective operators that arise in a compositeness scenario. These are routinely used in experimental searches at the LHC to constraint contact and gauge interactions between ordinary Standard Model fermions and excited (composite) states of mass $M$.
After deducing the unitarity bound for the production process of a composite neutrino, we implement such bound   
and compare it with the recent experimental exclusion curves for Run 2, the High-Luminosity and High-Energy configurations of the LHC. Our results also applies to the searches where a generic single excited state is produced via contact interactions.
We find that the unitarity bound, so far overlooked, is quite compelling and significant portions of the parameter space ($M,\Lambda$) become excluded in addition to the standard request $M \le \Lambda$. 
\end{abstract}


\end{frontmatter}
\makeatletter
\def\ps@pprintTitle{%
   \let\@oddhead\@empty
   \let\@evenhead\@empty
   \let\@oddfoot\@empty
   \let\@evenfoot\@oddfoot
}
\makeatother

\section{Introduction}
\label{sec:Intro}
It is well known that partial wave unitarity is a powerful tool to estimate the perturbative validity of effective field theories (EFTs). It has been used in the past to provide useful insights both in strong and electroweak interactions~\cite{Lee:1977yc} as well as in quantum gravity~\cite{Atkins:2010re}. Perhaps the best known example is the  bound on the Higgs mass derived from an analysis of $WW\to WW$ scattering within the Standard Model  (SM)~\cite{Lee:1977yc,Lee:1977eg}. On the other end, unitairty has also been applied to a number of approaches  beyond the Standard Model (BSM). For instance in composite Higgs models~\cite{DeCurtis:2016scv}, in searches of scalar di-boson resonances~\cite{DiLuzio:2017qgm,Luzio2017xxx}, searches for dark matter effective interactions~\cite{Endo:2014mja} and on generic dimension-6 operators~\cite{Corbett:2017qgl}.   

One possible BSM alternative, widely discusses in literature and routinely pursued in high-energy experiments, is a    
composite-fermions scenario which offers a possible solution to the hierarchy pattern of fermion masses~\cite{Pati:1975md,Pati:1983dh,Harari:1982xy,Greenberg:1974qb,Dirac:1963aa}.
In this context~\cite{Terazawa:1976xx,Eichten:1979ah,Eichten:1983hw,Cabibbo:1983bk,Baur:1989kv,Baur:1987ga}, SM quarks ``$q$'' and leptons ``$\ell$'' are assumed to be bound states of some as yet not observed fundamental constituents generically referred 
as {\it preons}. If quarks and leptons have an internal substructure, they are expected to be accompained by heavy excited states $\ell^*, q^*$ of masses $M$ 
that should manifest themselves at an unknown energy scale, the compositeness scale 
$\Lambda$.

As customary in an EFT approach, the effects of the high-energy physics scale, here $\Lambda$, are captured in higher dimensional operators that describe processes within a lower energy domain, where the fundamental building blocks of the theory cannot show up. Hence, the heavy excited states may interact with the SM ordinary fermions via dimension-5 gauge interactions of the SU(2)$_L \otimes$ U(1)$_Y$ SM gauge group of the magnetic-moment type (so that the electromagnetic current conservation is not spoiled by e.g.~$\ell^* \ell \gamma$ processes~\cite{Cabibbo:1983bk}). In addition, the exchange of preons and/or 
binding quanta of the unknown interactions between ordinary fermions ($f$) and/or the excited states ($f^*$) results in effective
contact interactions (CI) that couple the  SM fermions and heavy excited states~\cite{Baur:1989kv,Baur:1987ga,Peskin:1985symp,Tanabashi:2018oca}. 
In the latter case, the dominant effect is expected to be given by the dimension-6 four-fermion  interactions  scaling with the inverse square of the compositeness scale $\Lambda$:
\begin{subequations}
\label{contact}
\begin{align}
\label{Lcontact}
\mathcal{L}^{6}&=\frac{g_\ast^2}{\Lambda^2}\frac{1}{2}j^\mu j_\mu \, , \\
\label{Jcontact}
j_\mu&=\eta_L\bar{f_L}\gamma_\mu f_L+\eta\prime_L\bar{f^\ast_L}\gamma_\mu f^\ast_L+\eta\prime\prime_L\bar{f^\ast_L}\gamma_\mu f_L + h.c.  \nonumber\\&\phantom{=} +(L\rightarrow R) \, ,
\end{align}
\end{subequations}
where $g_*^2 = 4\pi$ and the $\eta$'s factors are usually set equal to unity. In this work the right-handed currents will be neglected for simplicity (this is also the setting adopted by the experimental collaborations).

As far as gauge interactions  (GI) are concerned, 
let us consider the first lepton family and assume that the excited neutrino and the excited electron are grouped into left handed singlets and  a right-handed SU(2) doublet:
$
e^*_L,  \nu_L^*\, , 
L^*_R=\left(
 \nu^*_R \, , e^*_R 
\right)^T
$,
so that  a magnetic type coupling between the left-handed SM doublet and the right-handed excited doublet via the SU(2)$_L \otimes$ U(1)$_Y$ gauge fields can be written down~\cite{Cabibbo:1983bk,Takasugi:1995bb}:
\begin{equation}
\label{mirror}
{\cal L}^5 =\frac{1}{2 \Lambda} \, \bar{L}_R^* \sigma^{\mu\nu}\left( gf \frac{\bm{\tau}}{2}\cdot \bm{W}_{\mu\nu} +g'f' Y B_{\mu\nu}\right) L_L +h.c. \, .
\end{equation}
Here, $L^T =({\nu_\ell}_L, \ell_L)$ is the ordinary lepton doublet, $g$ and $g'$ are the SU(2)$_L$ and U(1)$_Y$ gauge couplings and $\bm{W}_{\mu\nu}$, $B_{\mu\nu}$ are the field strength tensor of the corresponding gauge fields respectively; $\bm{\tau}$ are the Pauli matrices and $Y$ is the hypercharge,  $f$ and $f'$ are dimensionless couplings and are expected (and assumed) to be of order unity. 
\begin{figure}[t!]
\centering
\includegraphics[scale=0.3175]{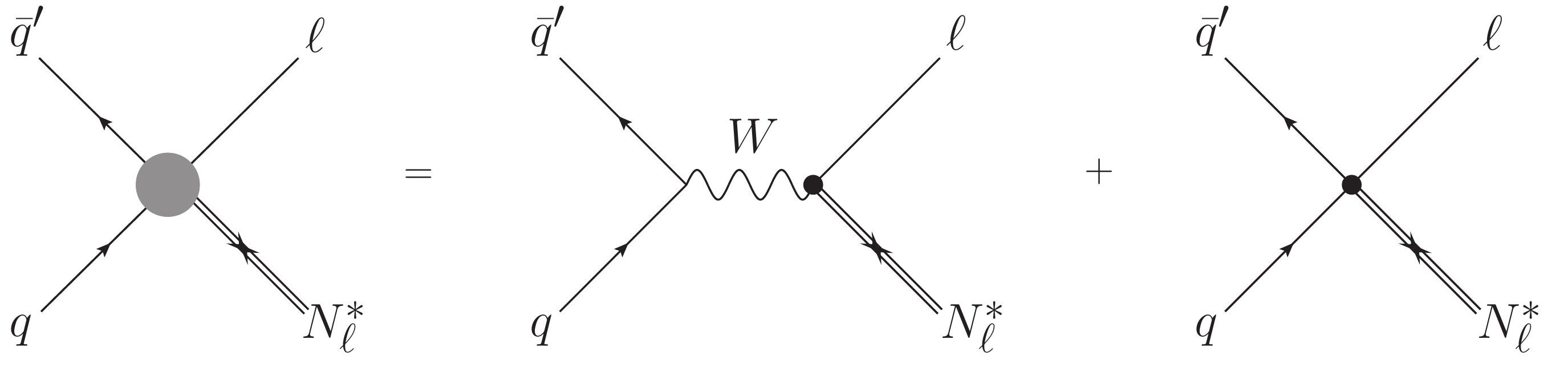}
\caption{\label{fig:process}Feynman diagrams depicting the mechanisms responsible for the process $q \bar{q}' \to N^* \ell$, where $\ell$ stands for both $\ell^\pm$. The dark grey blob (diagram on the left) describes the production of an on-shell heavy Majorana neutrino $N$ in proton-proton collisions at
LHC. The production is possible both with gauge interactions (first diagram on the right-hand side) and with four-fermion contact
interactions (second diagram on the right-hand side).}
\end{figure}

Excited states interacting with the SM sector through the model Lagrangians (\ref{Lcontact})-(\ref{Jcontact}) and (\ref{mirror}) have been extensively searched for at high-energy collider facilities. The current  strongest bounds are due to the recent LHC experiments. Charged leptons ($e^*
,\mu^*$) have been searched for in the channel $pp\to \ell \ell^* \to \ell \ell \gamma$~\cite{atlas-limit,atlas-limit-new,cms-limit-7TeV,cms-limit-8TeV,cms-limit-13TeV,CMS-PAS-EXO-18-004,Sirunyan:2018zzr}, i.e. produced via CI and then decay via GI, and in the channel $pp \to \ell \ell^* \to \ell \ell q\bar{q}^\prime$  \cite{CMS-PAS-EXO-18-013} where both production and decay proceed through CI.
Neutral excited leptons have been also discussed in the literature and the corresponding phenomenology at LHC has been discussed in detail in the case of a heavy composite Majorana neutrino $N^*$~\cite{Leonardi:2015qna}. 
A dedicated experimental analysis has been carried out 
by the CMS collaboration \cite{Sirunyan:2017xnz} on LHC data collected for $\sqrt{s}$ = 13 TeV and looking for the process \begin{equation}pp\rightarrow \ell N^*_\ell \,  \rightarrow \ell\ell q\bar{q}^\prime\end{equation}
with dilepton (dielectrons or dimuons) plus diquark final states.  The existence of $N^*_\ell$ is excluded for masses up to $4.60$ $(4.70)$ TeV at $95\%$ confidence level, assuming $M = \Lambda$. Moreover, the composite Majorana neutrinos of this model can be responsible for baryogenesis via leptogenesis~\cite{Zhuridov:2016xls,Biondini:2017fut}.
The phenomenology of other excited states has also been discussed in a series of recent papers~\cite{Biondini:2012ny,Majhi:2012bx,Leonardi:2014epa,Biondini:2014dfa,Leonardi:2015qna,Panella:2017spx,Presilla:2018ryu,Caliskan:2018vsk}.

We emphasize that in all phenomenological studies referenced above as well as all experimental analyses that have searched for excited states at colliders, it is customary to impose the constraint $M \le \Lambda$ on the parameter space of the model. To the best of our knowledge unitarity has never been taken into account and/or discussed in connection with the effective interactions of the so called excited states.
The main goal of this work is to report instead that the unitarity bounds, as extracted from Eq.~(\ref{Lcontact})-(\ref{Jcontact}) and (\ref{mirror}), are quite compelling and should be included in future studies of such effective composite models because they contraint rather strongly the parameter space.  
While we present an explicit calculation of the unitarity bound for heavy composite neutrino searches, we expect that similar bounds (i.e. equally compelling) would apply for excited electrons ($e^*$), muons ($\mu^*$) and quarks ($q^*$). Indeed, the effective operators that describe the latter excited states have the very same structure of those referred to the composite neutrinos.

\section{Unitarity in single-excited-fermion production}
\label{unitarity}
For the derivation of the unitarity bound, we adopt a standard method that makes use of the optical theorem and the expansion of the scattering amplitude in partial waves. In order to specify the CI and GI Lagrangians for a definite situation, we consider the production of the excited Majorana neutrino at the LHC. However, we shall highlight when the results apply to other composite fermion states in the following.

The central object that we shall derive from the operators in the effective Lagrangians (\ref{CI_Lag_LH}) and (\ref{GI_Lag_LH}) is the interacting part of the $S$ matrix, indicated with $T$ in this letter. It enters the partial wave decomposition of the scattering amplitude as follows 
\begin{equation}
\mathcal{M}_{i \to f} (\theta ) = 8 \pi \sum_j (2 j +1 ) T^j_{i \to f} d^j_{\lambda_f \, \lambda_i} (\theta) \, ,
\label{PW_dec}
\end{equation}
where $j$ is the eigenvalue of the total angular momentum $J$ of the incoming (outgoing) pair,  $d^j_{\lambda_f \, \lambda_i} (\theta)$ is the Wigener d-function and $\lambda_i$ ($\lambda_f$) is the total helicity of the initial (final) state pair. Without loss of generality, we consider azimuthally symmetric processes and fix $\phi=0$ accordingly.  From the optical theorem and the decomposition in Eq.~(\ref{PW_dec}), one can find the perturabtive unitarity condition of an inelastic process for each $j$ to be
\begin{equation}
\sum_{f \neq i} \beta_i \beta_f |T^j_{i \to f}|^2 \leq 1 \, ,
\label{Bound_Inelastic}
\end{equation}
where $\beta_i$ ($\beta_f$) is the factor obtained from the two-body phase space and reads for two generic particles with masses $m_1$ and $m_2$
\begin{equation}
\beta = \frac{\sqrt{\left[ \hat{s}-(m_1-m_2)^2\right] \left[ \hat{s}-(m_1+m_2)^2\right]}}{\hat{s}} \, .
\label{Beta_fi}
\end{equation}
It corresponds to the particle velocity when $m_1=m_2$. It is important to notice that the unitarity bound is imposed on the subprocess involving the proton valence quarks as initial state, namely $q \bar{q}' \to \ell N^*_\ell$ as shown in Figure~\ref{fig:process}. Then, for the process of interest, the relevant interaction(s) are as follows: 
    \begin{equation}
    \mathcal{L}_{\text{CI}}= \frac{g_*^2 \, \eta}{\Lambda^2} \bar{q}' \gamma^\mu P_L q \, \bar{N} \gamma_\mu P_L \ell  + h.c. \, ,
    \label{CI_Lag_LH}
\end{equation}

\begin{equation}
    \mathcal{L}_{\text{GI}}= \frac{g \, f}{\sqrt{2} \Lambda} \bar{N} \sigma^{\mu \nu} (\partial_\mu W_\nu^+) P_L \ell + h.c. \, .
    \label{GI_Lag_LH}
\end{equation}
Accordingly, in Eq.~(\ref{Beta_fi}), $\hat{s}$ denotes the center-of-mass energy in each collision and it is obtained from the nominal collider energy and the parton momentum fractions as $\hat{s}= x_1 x_2 \, s$.  As far as the kinematic is concerned, $\beta_i=1$ can be used since the valence quark masses are negligible with respect to the center-of-mass energy. Instead, one finds  
$\beta_f = 1-M^{2}/\hat{s}$ for the final state, where the composite neutrino mass has to be kept. 

The core of the method relies on the derivation of the amplitude for the process of interest induced by the contact and gauge-mediated effective Lagrangians (\ref{CI_Lag_LH}) and (\ref{GI_Lag_LH}). Then, one matches the so-obtained result for $\mathcal{M}_{i \to f}$ with the r.h.s of Eq.~(\ref{PW_dec}) and extracts the corresponding $T^j_{i \to f}$ for each definite eigenvalue of the total angular momentum ($j$). The latter are inserted into Eq.~(\ref{Bound_Inelastic}) in order to derive the unitarity condition that the model parameters ($\Lambda, M, g_*,g$) and the center-of-mass energy have to obey.  To this end, the amplitude $\mathcal{M}_{i \to f}$ is decomposed in terms of definite helicity states and, therefore, we have to express the initial and final state particles spinors accordingly \cite{Jacob:1959at}. The helicity of each particle in the initial or final state is $\lambda=\pm 1/2$, being all the involved particles fermions (also the composite neutrinos are spin-1/2 fermion) . We shall simply use $\pm$ to label the initial and final state helicity combinations, $(+,+)$, $(+,-)$, $(-,+)$ and $(-,-)$. Since in the center-of-mass frame the incoming and outgoing particles travel in opposite directions, the helicities in the Wigner d-functions are defined as $\lambda_i=\lambda_q - \lambda_{\bar{q}'}$ and $\lambda_f=\lambda_{N^*} - \lambda_{\ell}$.  One can adopt two different bases for expressing the spinors and the gamma matrices, the Dirac and chiaral bases (see e.g. appendix in ref.~\citep{Endo:2014mja}). We used both the options to derive $T^j_{i \to f}$ and checked that our findings are indeed  invariant upon the choice of the basis.

We give the result for the CI Lagrangian in Eq.~(\ref{CI_Lag_LH}) first. The non-vanishing helicity amplitudes read  
    \begin{eqnarray}
    && T^{j=1}_{(-,+) \to (-,+)}  = -\frac{\hat{s} \, g_*^2}{12 \pi \Lambda^2} \left( 1-\frac{M^{2}}{\hat{s}} \right)^{\frac{1}{2}} \, ,
\label{CI_Tampl_1}
    \\
    && T^{j=1}_{(-,+) \to (+,+)} =   \frac{\sqrt{\hat{s}} \, M \, g_*^2}{12 \sqrt{2}\pi \Lambda^2} \left( 1-\frac{M^{2}}{\hat{s}} \right)^{\frac{1}{2}} \, .
\label{CI_Tampl_2}
\end{eqnarray}
Only the amplitude with $j=1$ is non-zero, due to the initial helicity state. The same occurs with the vector and axial-vector operators studied in \citep{Endo:2014mja} for dark matter pair production at colliders. We notice that a finite composite neutrino mass allows for the helicity flip in the final state originating the term in Eq.~(\ref{CI_Tampl_2}). We obtain the same result if we work with right-handed particles in the CI operator, however the helicities in Eqs.~(\ref{CI_Tampl_1}) and (\ref{CI_Tampl_2}) flip as $+ \leftrightarrow -$.  Using Eq.~(\ref{PW_dec}) and summing over the non-vanishing final helicity states, we obtain
\begin{eqnarray}
    \frac{g_*^4 \, \hat{s} \, (2\hat{s}+M^{2})}{288 \pi^2 \Lambda^4} \left( 1-\frac{M^{2}}{\hat{s}} \right)^{2} \leq 1 \, .
    \label{UNI_CI}
\end{eqnarray}

As far as the GI process is concerned, we proceed the same way. A dimension-5 operator is involved and, in this case, the $W$ boson mediates the scattering between the initial and final states. We keep the $W$ boson mass in our expression, even if it is much smaller than the typical $\hat{s}$ values of the $pp$ collisions. The SM electroweak current enters besides the one from the composite model and the helicity amplitudes are found to be
\begin{eqnarray}
    && T^{j=1}_{(-,+) \to (-,+)}  = -\frac{i g^2}{24 \pi \Lambda} \frac{\hat{s}^{3/2}}{\hat{s}-m_W^2}\left( 1-\frac{M^{2}}{s} \right)^{\frac{1}{2}} \, ,
    \\
    && T^{j=1}_{(-,+) \to (+,+)} =   \frac{i g^2}{24 \sqrt{2}\pi \Lambda} \frac{\hat{s} \, M}{\hat{s}-m_W^2} \left( 1-\frac{M^{2}}{\hat{s}} \right)^{\frac{1}{2}} \, ,
    \end{eqnarray}
    and the corresponding result for the unitarity bound is 
    \begin{eqnarray}
    \frac{g^4 }{1152 \, \pi^2 \Lambda^2} \frac{\hat{s}^2 \, (2\hat{s}+M^{2})}{(\hat{s}-m_W^2)^2}  \left( 1-\frac{M^{2}}{\hat{s}} \right)^{2} \leq 1 \, .
    \label{UNI_GI}
\end{eqnarray}
A comment is in order. The unitarity bound in Eq.~(\ref{UNI_CI}) is valid for the more generic production process $q \bar{q}' \to f^* f$, i.e. excited charged or neutral leptons and excited quarks accompanied by a SM fermion. This statement traces back to the particle-blind choice adopted in the CIs framework, where the $\eta$'s are set to unity in all the cases. More care has to be taken about a wider applicability of the result for GIs in Eq.~(\ref{UNI_GI}). Here, different factors can enter according to the gauge couplings and gauge bosons that describe the processes involving excited charged leptons and quarks instead of composite neutrinos.  

\section{Implementing the bound}
\label{scalarUNC}
The production of heavy composite Majorana neutrinos has been studied by the CMS Collaboration by studying the final state with two leptons and at least one large-radius jet, with data from $pp$ collisions at $\sqrt{s} = 13$ TeV and with an integrated luminosity of $2.3$ fb$^{-1}$~\cite{Sirunyan:2017xnz}. Good agreement between the data and the SM expectations was observed in the search, but the whole dataset of the Run 2 of the LHC still needs to be analysed. Therefore, the issue of the unitarity condition on the accessible parameter space ($M,\Lambda$) urges to be assessed. As usual in BSM searches, the absence of a signal excess over the SM background is translated into an experimental bound on the parameter space $( M,\Lambda)$. 
Moreover, the sensitivity of this search was investigated for two future collider scenarios: the High-Luminosity LHC (HL-LHC), with a centre-of-mass energy of 14 TeV and an integrated luminosity of 3 ab$^{-1}$,  and the High-Energy LHC (HE-LHC), with a centre-of-mass energy of 27 TeV and an integrated luminosity of 15 ab$^{-1}$~\cite{CMS-PAS-FTR-18-006}. The projection studies, included in the recent Yellow Report CERN publication~\cite{CidVidal:2018eel,Collaborations:2651134}, have shown the potential of such facilities in reaching much higher neutrino masses.

In this section, the  perturbative unitarity bounds are applied to these searches in the dilepton and a large-radius jet channel with the CMS detector for the three different collider scenarios.
As already clear from the rather
different coupling values entering the Lagrangians (\ref{CI_Lag_LH}) and (\ref{GI_Lag_LH}), namely $g/ \sqrt{2} \approx 0.4$ versus $g_*^2=4\pi$,  the production mechanism of a heavy composite neutrino and other excited states is dominated by the contact interaction mechanism~\cite{Panella:2017spx}. In particular, it was shown that cross sections in contact-mediated production are usually more than two orders of magnitude larger than the gauge mediated ones for all values of the $\Lambda$ and $M$ relevant in the analyses. This means that it is a reasonable approximation to consider only the bounds given in Eq.~(\ref{UNI_CI}) to constraint the unitarity violation of the signal samples.

In order to estimate the effect of the unitarity condition on LHC searches, we need to implement the bounds  in the case of hadron collisions. Then, the square of the centre-of-mass energy of the colliding partons system, $\hat{s}=x_1 x_2 s$ does not have a definite value, where $x_1$ and $x_2$ are the parton momentum fractions and $\sqrt{s}$ is nominal energy of the colliding protons. To this aim, we have estimated $\hat{s}$ in each event generated in the Monte Carlo (MC) samples, and we have plugged the result into Eq.~(\ref{UNI_CI}) in order to obtain level curves on the parameter space for which the unitarity bound is satisfied to some extent.
Indeed, the constraint in Eq.~(\ref{UNI_CI}) should not be interpreted too strictly. A violation of such bounds would signal the breakdown of the EFT expansion and call for higher order operators in $\sqrt{\hat{s}}/ \Lambda$ to help in restoring the unitarity of the process. Therefore, we implement a theoretical uncertainty by allowing up to 50\% of the events to violate the bound, that corresponds to assigning a relative correction to the cross section $\delta \sigma / \sigma_{\hbox{\tiny LO}} \leq 0.5$ from higher order terms.

The MC samples for the signal are generated at Leading Order (LO) with CalcHEP (v3.6)~\cite{Belyaev:2012qa} for $\sqrt{s}=13, 14$ and $27$ TeV proton-proton collisions, using the NNPDF3.0 LO parton distribution functions with the four-flavor scheme~\cite{Ball:2014uwa}, spanning over the $(\Lambda,M)$ region covered by the experimental searches~\cite{CMS-PAS-FTR-18-006}.
The information on the parton momenta is then retrieved from the Les Houches Event (LHE) files of each signal process through MadAnalysis~\cite{Conte:2012fm}. 
We have explicitly checked that, in the mass range explored, the $\hat{s}$-distributions in our MC simulations are peaked at values around $M^{2}$ almost irrespective of the nominal collider energy $\sqrt{s}=13,14,27$ TeV. This is somehow expected on general grounds since in the generated signal events the available energy, $\sqrt{\hat{s}}$, is mostly used to produce a heavy excited state (of mass $M$). Of course, the larger the collider energy the more prominent the distribution tails at high $\hat{s}$ values. These expectations are corroborated by analytical expressions for the $\hat{s}$-distributions that can be retrieved from ref.~\cite{Leonardi:2015qna}, involving only the product of the parton luminosity functions and the hard production process cross section. 
 
\begin{figure}[t!]
\includegraphics[width=0.48\textwidth]{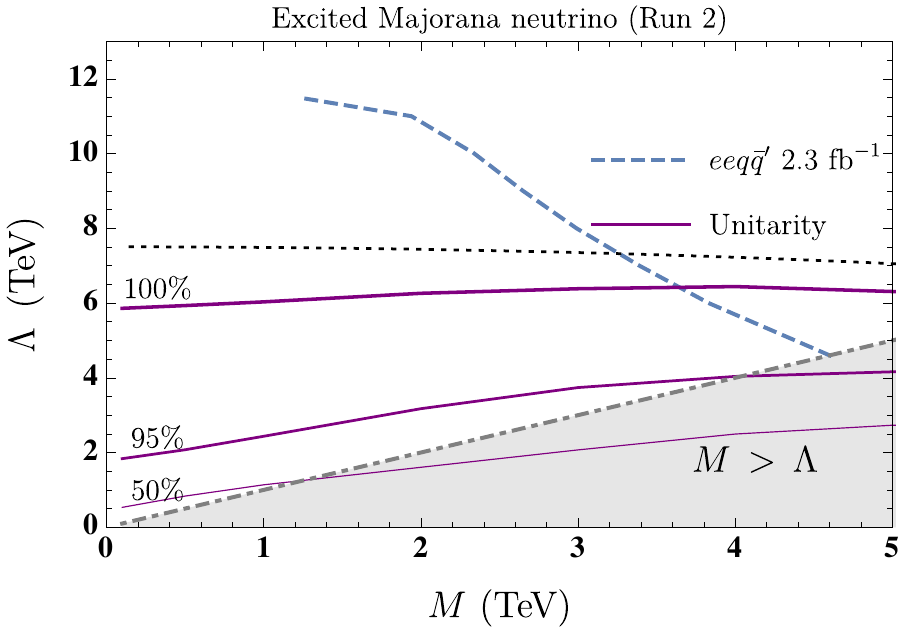} 
\caption{
\label{figex2} 
(Color online) The unitarity bound in the ($M,\Lambda$) plane compared with the Run 2 exclusion at 95\% CL from \protect\cite{Sirunyan:2017xnz}, dashed line (blue), for the $eeq\bar{q}'$ final state signature. The solid (violet) lines with decreasing thickness represent the unitarity bound respectively for 100\%, 95\% and 50\% event fraction satisfying Eq.~\ref{UNI_CI}. The dot-dashed (gray) line stands for the $M= \Lambda$ condition. Here and in the following figures both $\Lambda$ and $M$ start at 100 GeV, and the dotted (black) curve corresponds to the theoretical unitarity bound (Eq.~\ref{UNI_CI} with $\hat{s}=s$).
}
\end{figure}

\begin{figure}[t!]
\includegraphics[width=0.48\textwidth]{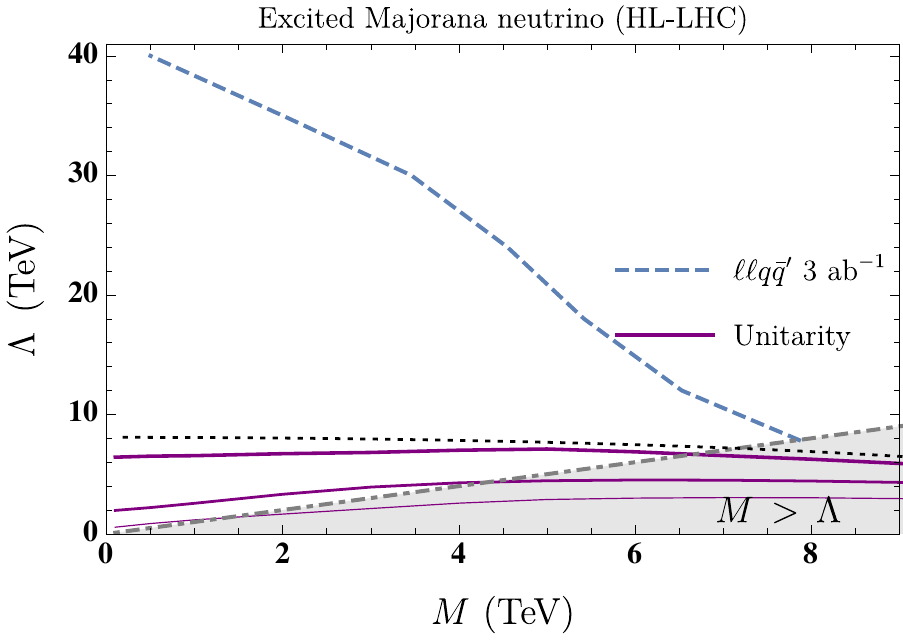}  
\caption{\label{figex3} (Color online)
The unitarity bound in the plane $(M, \Lambda) $ for the three event fractions as in Fig.~\ref{figex2} compared with the exclusion from the High Luminosity projections study in \cite{CidVidal:2018eel} for LHC at $\sqrt{s}=14$ TeV at 3 ab$^{-1}$ of integrated luminosity.}
\end{figure}

\begin{figure}[t!]
\includegraphics[width=0.48\textwidth]{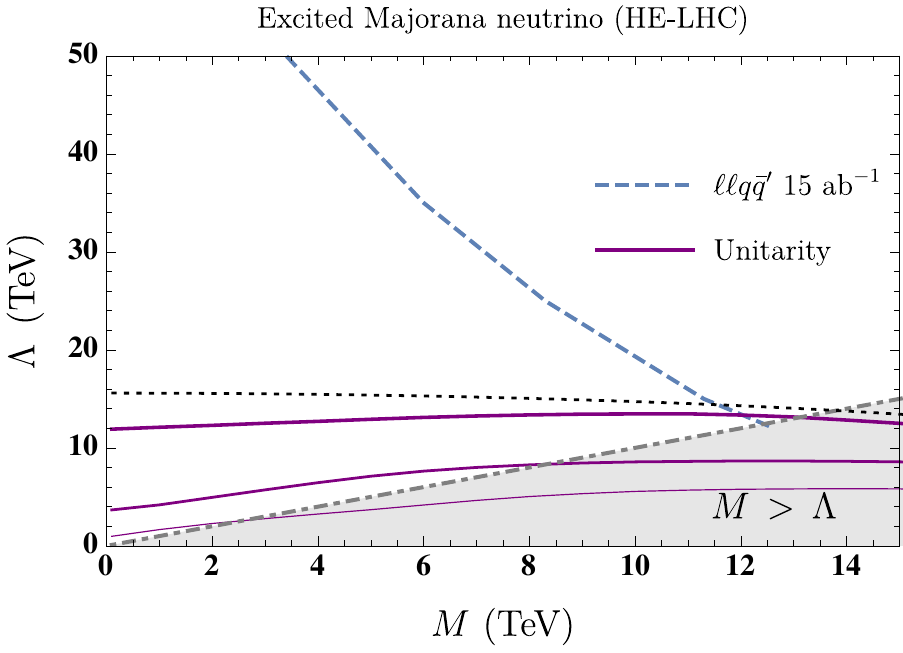}  
\caption{\label{figex4} (Color online)
 The unitarity bound in the plane $(M, \Lambda) $ for the three event fractions as in Fig.~\ref{figex2} compared with the exclusion curve from the HE-LHC projection studies in \protect\cite{CidVidal:2018eel} for $\sqrt{s}=27$ TeV at 15 ab$^{-1}$ of integrated luminosity.}
\end{figure}

\begin{figure}[htb!]
\includegraphics[width=0.48\textwidth]{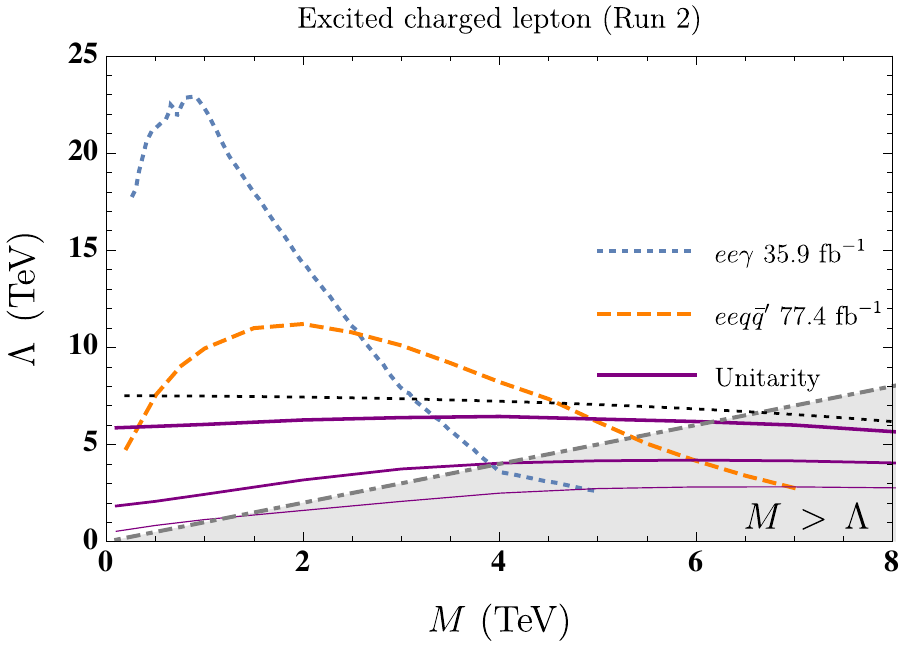}  
\caption{\label{figex5} (Color online)
The unitarity bound in the plane $(M, \Lambda) $ for the three event fractions as in Fig.~\ref{figex2} compared with the exclusion from the Run 2 for charged leptons searches with two different final states~\cite{Sirunyan:2018zzr,CMS-PAS-EXO-18-013}.}
\end{figure}

The results are presented in  Figs.~\ref{figex2},~\ref{figex3} and~\ref{figex4} for the Run 2, HL-LHC and HE-LHC scenario respectively.
The regions below the solid (violet) lines in Figs.~\ref{figex2}-\ref{figex4} delimit the parameter space for which respectively 100\%, 95\% and 50\% of the events satisfy the unitarity bound, i.e. they define the regions where the model should not be trusted because unitarity is violated for such $(M, \Lambda)$ values. It is important to underline that the impact of the unitarity bound is strongly dependent on this fraction of events ($f$) that satisfy the condition in Eq.~(\ref{UNI_CI}). This conforms with the results in~\cite{Endo:2014mja}, at least in the $(\Lambda, M)$ region considered here.
Moreover, we observe that there is a value of the compositeness scale $\Lambda$, which depends on the parton collision energy $\sqrt{\hat{s}}$ and the excited fermion mass $M$, above which the unitarity bound saturates. Indeed, one can estimate an upper bound for such a value from Eq.~(11) by setting the collision energy $\sqrt{\hat{s}}=\sqrt{s}$; it is represented with the dotted (black) line in Fig.~\ref{figex2}-\ref{figex5} for the corresponding nominal energies $\sqrt{s}=13, \, 14, \, 27$ TeV.
An approximated (maximal) value of $\Lambda \approx\sqrt{s/3}$, which saturates  the unitary bound, is obtained when $s \gg M^2 $.


\section{Discussion and Results}
\begin{table*}[htb!]
    \centering
    \begin{tabular}{||c|c|c|c||}
\hline\hline & & &\\
         & LHC Run 2 ($N^*$) & LHC Run 2 ($e^*$) & LHC Run 2 ($e^*$)   \\
         & 2.3 fb$^{-1}, \sqrt{s}=13$ TeV
         & 35.9 fb$^{-1}, \sqrt{s}=13$ TeV
         & 77.4 fb$^{-1}, \sqrt{s}=13$ TeV\\
         & & &\\\hline& & &\\
 $M =\Lambda$ & $M \le 4.6 $ TeV~\protect\cite{Sirunyan:2017xnz}  & $M \le 4.0 $ TeV ~\protect\cite{ Sirunyan:2018zzr}& $M \le 5.5 $ TeV~\protect\cite{CMS-PAS-EXO-18-013}\\ & & &\\
 Unitarity 100\%  & $M \le 3.6$ TeV \,\,\,($\Lambda = 6.4  $ TeV) & $M \le 3.3$ TeV\,\,\,  ($\Lambda = 6.5 $ TeV) &$M \le 4.9$ TeV \,\,\,  ($\Lambda = 6.4  $ TeV) \\ & & &\\
 Unitarity 50\%  & $-$ & $M \le 4.9$ TeV\,\,\,  ($\Lambda = 2.8$ TeV) &$M \le 7.0$ TeV \,\,\,  ($\Lambda = 2.9 $ TeV) \\ & & &\\
\hline\hline
    \end{tabular}
    \caption{In the first line we quote the bounds reported in the CMS analysis of like sign dilpetons and diquark for excited neutrinos \protect\cite{Sirunyan:2017xnz} and the bounds from CMS for two analyeses for excited charged leptons~\cite{Sirunyan:2018zzr,CMS-PAS-EXO-18-013}. In second (third) line, we quote instead the strongest mass bound obtained from Figs.~\ref{figex2} and \ref{figex5} when the  perturbative unitarity bound with $f= 100\%$ (50\%)   crosses the 95\% C.L. exclusion curve from the experimental studies.    }
    \label{tab:I}
\end{table*}

Let us elaborate on our findings and explain their impact on the experimental analyses carried out at the LHC. First of all, the experimental outcomes are summarized with exclusion regions in the $(M,\Lambda)$ plane, which are in turn set with the 95\% C.L. observed (Run 2)~\cite{Sirunyan:2017xnz} and expected limit (HL/HE-LHC)~\cite{CidVidal:2018eel,Collaborations:2651134}, namely the dashed blue lines in Figure~\ref{figex2}, \ref{figex3} and \ref{figex4} respectively. Above these lines the model is still viable, whereas below it is excluded. The experimental collaborations quote routinely the largest excluded excited-state mass by intersecting the 95\% C.L. exclusion curves with the $M \leq \Lambda$ constraint (dot-dashed gray line and gray shaded region in Figure~\ref{figex2}, \ref{figex3} and \ref{figex4}). This is the widely adopted condition imposed on the model validity and it originates from asking the heavy excited states to be at most as heavy as the new physics scale $\Lambda$. Despite it is a reasonable constraint, it does not take into account the typical energy scale that enters the production process, i.e.~$\sqrt{\hat{s}}$.

Comparing the unitarity bounds, as represented by the solid violet lines in Figure~\ref{figex2}, \ref{figex3} and \ref{figex4}, with the usual prescription of $M \leq \Lambda$ we can frame the interplay between the two. For small values of the heavy neutrino masses ($M$), the unitarity bound is more restrictive than $M \leq \Lambda$ and it shrinks the available parameter space quite considerably (at least for $f=100\%$ and $95\%$). On the other hand, for higher values of the masses the two becomes compelling. The relative importance depends on both the collider nominal energy and event fraction $f$. 
If one applies the unitarity bound to the experimental results by following the same prescription as outlined before for $M \leq \Lambda$, then the maximal neutrino mass values are those collected in Table~\ref{tab:I}, second raw. For example, for LHC Run 2, we find $M \le  3.6$ TeV for $\Lambda= 6.4 $ TeV instead of $M \le  4.6$ TeV ($\Lambda=4.6$ TeV), when the unitarity bound is required to be satisfied by the 100\% of events. As anticipated, the new constraint set by the unitarity bound offers an alternative theoretical input for  ongoing and future experimental analyses on this effective composite model. We provide the unitarity bound down to $M=100$ GeV. Smaller values clash with the original model setting that assumes some new physics above the electroweak scale triggering fermion excitations~\cite{Baur:1989kv,Cabibbo:1983bk}.

While in this study we have concentrated on the impact of unitarity bounds on the heavy composite neutrino production at the LHC, HL-LHC and HE-LHC, we expect that similar bounds will affect the searches for charged excited leptons. Leaving more detailed studies for future work, we apply the  unitarity bound in Eq.~(\ref{UNI_CI}) for the recent experimental results reported in~\cite{Sirunyan:2018zzr,CMS-PAS-EXO-18-013}, where charged excited leptons are produced via CI in the process $pp \to \ell^* \ell$, with $\ell^*=e^*,\mu^*$. Being the effective Lagrangian for the production mechanism the very same as for excited Majorana neutrinos (if one insists on $\eta=1$ for the different states), we can use the unitarity bound as extracted for $\sqrt{s}=13$ TeV and apply it for the process involving charged excited leptons. The comparison with the experimental exclusion limit at 95\% C.L. is shown in Figure~\ref{figex5}.     
As anticipated, the comparison with the observed and expected limits produces different mass reaches. For the case of LHC Run 2 searches recently performed in CMS, we quote the corresponding mass values in Table~\ref{tab:I} for the searches of excited charged leptons. As for the analyses on the excited charged states, we can exploit the data as provided in the region $M>\Lambda$ and inspect the interplay with the unitarity bound for $f=100\%,95\%,50\%$. On the other hand, we cannot provide as many mass values for the excited neutrino searches, due to the lack of experimental data in the same region $M>\Lambda$. When $f=100\%$ is considered, the largest excited lepton mass reads 3.3 TeV (4.9 TeV) instead of 4.0 TeV (5.5 TeV) for the $ee\gamma$ ($eeq\bar{q}'$) final state. 

\emph{In conclusion}, we studied the perturbative unitarity bound extracted from the effective gauge and contact Lagrangians for a composite-fermion model. On general grounds, an effective theory is valid up to energy/momentum scales smaller than the large energy scale that sets the operator expansion. Since collider experiments are involving more and more energetic particle collisions, the use and the applicability of effective operators can be questioned. In order to address this issue and to be on the safe side, one can impose the unitarity condition both on the EFT parameters $(M, \Lambda, g^*, g)$ and the energy involved in a given process. To the best of our knowledge, such a constraint was not derived for the model Lagrangians in Eq.~(\ref{CI_Lag_LH}) and (\ref{GI_Lag_LH}), and we have obtained the corresponding unitarity bounds, namely Eqs.~(\ref{UNI_CI}) and (\ref{UNI_GI}). Thus, the applicability of the effective operators describing the production of composite neutrinos (and other excited states) has to be restricted accordingly. We have considered an estimation of the theoretical error on the unitarity bound, which is derived at leading order in the EFT expansion, by allowing up to 50\% of the events to evade the constraint. This originates the violet lines in Figures~\ref{figex2}-\ref{figex5}.

On the basis of the results, further investigations may be devoted to a better understanding of the theoretical error. Indeed the strong dependence of the unitarity bound on the fraction of events $f$, especially so in the low-mass region, calls perhaps for an estimate of possible higher order terms in the effective theory expansion (operators of dimension-7 for contact inteactions). In doing so, one could pinpoint to a particular choice of $f$ in a more rigorous way.

Nonetheless, it is the authors' opinion that the findings here discussed will have a significant impact on ongoing and future experimental searches for excited states coupling to the SM fermions with the interactions given in (\ref{CI_Lag_LH}) and (\ref{GI_Lag_LH}). At the very least, the unitarity bounds play the role of a collider-driven theoretical tool 
for interpreting the experimental results of the considered composite models,
more rigorous  
than the simple relation $M \leq \Lambda$.
\section*{Acknowledgments} 
The authors thank P.~Azzi and F.~Romeo for useful comments and discussion on the manuscript. We are especially thankful to Dr. Oleg Zenin and Dr. Andrey Kamenshchikov (ATLAS Collaboration) for pointing out an inconsistency between our  previous numerical results and the theoretical formula of the unitarity bound, and for crosschecking some of our revised numerical results.




\bibliographystyle{elsarticle-num} 
\bibliography{unitarity.bib}





\end{document}